# On the Amplitude of Vortex Entropy: A Semiclassical Treatment

Yehao Guo,[1,†] Dong Qiu,[1] Haiwen Liu,[2] and Jie Xiong[1,*]

[1]*School of Physics and State Key Laboratory of Electronic Thin Films and Integrated Devices,*

*University of Electronic Science and Technology of China, Chengdu 611731, China*

[2]*Department of Physics and Center for Advanced Quantum Studies, Beijing Normal University,*

*Beijing 100875, China*

Despite a long history of Nernst effect in superconductors, a satisfactory theory on its amplitude in vortex liquid phase is still absent. The central quantity of vortex Nernst signals is the entropy $s_\phi$ carried by each vortex. Here we show a semiclassical treatment based on London equation and Pippard nonlocal generalization. The derived $s_\phi$ is a function of both temperatures and magnetic fields. Its magnitude $s_\phi^{\mathrm{amp}}$ scales with normal-state conductivity $\sigma_\mathrm{n}$. Estimations based on our formula show good consistency with experimentally determined values. In dirty limit, the relation is further simplified into a Wiedemann-Franz-like ratio $s_\phi^{\mathrm{amp}}/\sigma_\mathrm{n} \sim k_\mathrm{B} \ln 2 /\sigma_\mathrm{Q}$ if taking parameter values deduced from Homes' law, where $\sigma_\mathrm{Q} = 4e^2/h$ is two-dimensional quantum conductivity. We also address related issues, including bounds to $s_\phi$, the Nernst signal and viscosity-entropy density ratio, which are all expressed in fundamental physical constants.

[†]Contact author: yehao.guo@std.uestc.edu.cn
[*]Contact author: jiexiong@uestc.edu.cn

***Introduction***. In the presence of magnetic fields, the Nernst effect is the transverse electric field under longitudinal thermal gradient. The Nernst signals can be generated either by quasiparticles, or by short-lived Cooper pairs or by vortices [1–3]. While the amplitude of the former two has been understood [1,2], there is no satisfactory account of the vortex Nernst effect. Vortices, each carrying a flux quantum $\Phi_0 = h/2e$, enter type-II superconductors under magnetic fields above the lower critical field $H_{c1}$. Its Nernst signal, experimentally defined as the ratio of transverse electric field to longitudinal temperature gradient, can be calculated through the relation $N = s_\phi \rho_{xx}/\Phi_0$, where $\rho_{xx}$ is the longitudinal resistivity and $s_\phi$ is the entropy carried by each vortex [4]. Compared with the more familiar $\rho_{xx}$, the central quantity is $s_\phi$. Stephen first investigated $s_\phi$ near $H_{c1}$ [5]. Sergeev *et al* obtained a smaller $s_\phi$ by considering vortex cores only [6]. However, the predicted $s_\phi$ exceeded the experimental value in Nb-doped strontium titanate (Nb: $SrTiO_3$) by fifty times [7]. The application of Uemura's law [8] to $Bi_2Sr_2CaCu_2O_{8+x}$ (BSCCO) predicted a constant $s_\phi$ and failed to capture the evolution of its magnitude upon doping [9]. The critical clue was pointed out by Behnia that vortex entropy may depend on the normal-state resistivity $\rho_n$ [3]. Historically, similar situation was encountered in 1950s by Pippard, who found the penetration depth $\lambda$ varied significantly with the addition of impurities, although $T_C$ was almost the same [10]. By generalizing London equation to a nonlocal form, confirmed later by Bardeen-Cooper-Schrieffer theory microscopically [11], Pippard connected $\lambda$ with the mean free path $l$ thus the effect of impurities, obtaining successful fitting to experimental data.

Inspired by Behnia and Pippard, we present a semiclassical theory on $s_\phi$ here. The free energy density of vortices $F_v$ is formulated based on London equation. The entropy density $S$ can then be derived through $S = -\partial F_v/\partial T$. Divided by areal density $n_v = B/\Phi_0$, we obtain the average entropy $s_\phi$ per vortex. The normal-state conductivity $\sigma_n$ is introduced by Pippard nonlocal generalization [10]. After the product of two dimensionless parameters is deduced empirically from Homes' law [12,13], we estimate the amplitude $s_\phi^{amp}$ of various superconductors, showing good consistency with experiments. In dirty limit, our result is simplified into $s_\phi^{amp}/\sigma_n \sim k_B \ln 2 /\sigma_Q$ with two-dimensional (2D) quantum conductivity $\sigma_Q = 4e^2/h$, reminiscent of Wiedemann-Franz law [14]. We also addressed bounds to $s_\phi^{amp}$, $N$ and viscosity-entropy density ratio $\eta/S$, which can all be expressed in fundamental physical constants. The minimum $s_\phi^{amp}$ to keep superconductivity in dirty limit is $k_B \ln 2$, and the upper bound to $N$ is $k_B \ln 2 /e$. The lower bound $\eta/S \geq \pi\hbar/k_B \ln 2$ resembles the holographic bound [15].

***Vortex entropy and its amplitude***. The free energy density $F_v$ of vortices consists of $F_{v1}$, gained outside vortex cores due to induction fields and kinetic energy of supercurrents, and $F_{v2}$ inside cores due to the loss of condensation. $F_{v1}$ is calculated by an integral in real space, or equivalently, summation in reciprocal space using Plancherel's theorem:

$$F_{v1} = \frac{1}{2\mu_0} \int dr \left[ \boldsymbol{h}(\boldsymbol{r})^2 + \lambda^2 (\nabla \times \boldsymbol{h}(\boldsymbol{r}))^2 \right] = \frac{1}{2\mu_0} \sum_{\boldsymbol{k}} (1 + \lambda^2 k^2) \boldsymbol{h}(\boldsymbol{k})^2 \quad (1)$$

$\boldsymbol{h}(\boldsymbol{k})$ is solved by Fourier transform of London equation $\nabla \times \nabla \times \lambda^2 \boldsymbol{h}(\boldsymbol{r}) + \boldsymbol{h}(\boldsymbol{r}) = \sum_i \Phi_0 \delta(\boldsymbol{r} - \boldsymbol{R}_i) \hat{\boldsymbol{z}}$ in a unit area:

$$h(\boldsymbol{k}) = \frac{n_v \Phi_0}{1 + \lambda^2 k^2} = \frac{B}{1 + \lambda^2 k^2} \quad (2)$$

$h(\boldsymbol{k})$ is nonzero only when $\boldsymbol{k}$ equals multiples of reciprocal lattice vector. For the sake of simplicity, we will treat it as a continuous variable, replacing the summation in Eq. (1) by an integral from $2\pi/a$ to $2\pi/2\xi$, where $a$ and $\xi$ are vortex lattice constant and coherence length, respectively. The constant 1 in the dominator of Eq. (2) can also be omitted, which is justified under fields high enough so that

$\lambda/\xi \geq \lambda k \geq \lambda/a \gg 1$. This procedure is equivalent to neglect the spatial variation of $h(\boldsymbol{r})$ except for a homogeneous background ($\boldsymbol{k} = 0$). The distribution of supercurrent, carrying kinetic energy around vortices, leads to a logarithmic dependence on the ratio $a/\xi$ or $\mu_0 H_{c2}/B$. The free energy density in Eq. (1) is thus expressed as:

$$F_{v1} = \frac{B^2}{2\mu_0} + \frac{n_v \Phi_0^2}{8\pi\mu_0 \lambda^2} \ln\frac{\mu_0 H_{c2}}{B} \tag{1'}$$

The loss of condensation inside the vortex core amounts to $\pi\xi^2\mu_0 H_c^2/2$ per vortex, where $H_c$ is the thermodynamic critical field with the relation $2\sqrt{2}\mu_0\pi\xi H_c \lambda = \Phi_0$ [16]. Therefore, the free energy density $F_{v2}$ is:

$$F_{v2} = \frac{n_v \Phi_0^2}{8\pi\mu_0 \lambda^2} \frac{1}{2} \tag{3}$$

Adding $F_{v1}$ and $F_{v2}$, we obtain the total free energy density of vortices:

$$F_v = \frac{B^2}{2\mu_0} + \frac{n_v \Phi_0^2}{8\pi\mu_0 \lambda^2}\left(\frac{1}{2} + \ln\frac{\mu_0 H_{c2}}{B}\right) \tag{4}$$

The temperature dependence is introduced through $\lambda$ and $H_{c2}$, by assuming:

$$\begin{aligned}\frac{1}{\lambda^2(T)} &= \frac{1}{\lambda^2(0)} f(\varepsilon) \\ H_{c2}(T) &= H_{c2}(0) f(\varepsilon)\end{aligned} \tag{5}$$

where $\varepsilon = 1 - T/T_C$, and $f(\varepsilon)$ is a non-decreasing function of $\varepsilon$ with $f(0) = 0$. The simplest form is Ginzburg-Landau (G-L) type $f(\varepsilon) = \varepsilon$ near $T_C$. The vortex entropy density is calculated through $S = -\partial F_v/\partial T$. Divided by $n_v$, we obtain the average entropy per vortex:

$$s_\phi(T, B) = \frac{\Phi_0^2}{8\pi\mu_0 \lambda^2(0) T_C} \frac{df}{d\varepsilon}\left(\frac{3}{2} + \ln\frac{\mu_0 H_{c2}}{B}\right) \tag{6}$$

$s_\phi$ is the function of both temperatures and magnetic fields, with its amplitude characterized by $s_\phi^{amp} = \Phi_0^2/8\pi\mu_0 \lambda^2(0) T_C$. According to G-L theory, $df/d\varepsilon = 1$ near $T_C$. As the temperature is lowered, $df/d\varepsilon$ will be smaller since $H_{c2}$ deviates from the G-L form and eventually saturates. The conclusion can also be arrived by analyzing the limit at $T = 0$. Starting from a superconducting ground state ($S = 0$) under zero field, the third law of thermodynamics requires $s_\phi$ is zero as the field is turned on, leading to $df/d\varepsilon = 0$ in Eq. (6). This implies $H_{c2}$ should saturate as $T$ approaching zero, consistent with the usual tendency of $H_{c2}(T)$ in experiments.

We utilize Pippard's nonlocal generalization [10] to $\lambda(0)$. The London penetration depth $\lambda_L$ is defined by $\lambda_L^2 = m/n_s e^2 \mu_0$, where $m$ and $e$ are electron mass and charge respectively, and $n_s$ is the density of superconducting electrons. When these quantities are those of Cooper pairs, $\lambda_L^2$ remain the same. We define two dimensionless parameters $\gamma = \hbar v_F/\xi k_B T_C$ and $\delta = n_s/n$, where $v_F$ is Fermi velocity and $n$ is the total electron density. Combining $l = v_F \tau$ and $\sigma_n = 1/\rho_n = ne^2\tau/m$, Pippard's $\lambda$ is expressed as:

$$\frac{1}{\lambda^2(0)} = \frac{1}{\lambda_L^2(0)} \frac{l}{\xi + l} = \delta\gamma\mu_0 \frac{k_B T_C}{\hbar} \frac{1}{\rho_n} \frac{1}{1 + l/\xi} \tag{7}$$

In dirty limit $l \ll \xi$, Eq. (7) is simplified into the same form as Homes' law [12,13] which gives $\delta\gamma \sim 2.8 \pm 0.6$ empirically:

$$\frac{1}{\lambda^2(0)} = \delta\gamma\mu_0 \frac{k_B T_C}{\hbar \rho_n} = \delta\gamma\mu_0 \frac{k_B}{\hbar} \sigma_n T_C \tag{7'}$$

The only difference is that $\rho_n$ here is taken at $T = 0$, while that of Homes' law is taken around $T_C$. With Eq. (7), the amplitude $s_\phi^{amp}$ can be rewritten as:

$$s_\phi^{\text{amp}} = \frac{\delta\gamma \Phi_0^2}{8\pi\rho_n} \frac{k_B}{\hbar} \frac{1}{1+l/\xi} = \frac{\delta\gamma}{4} \frac{k_B}{\sigma_Q} \frac{\sigma_n}{1+l/\xi} \quad (8)$$

where $\sigma_Q = 4e^2/h$. Eq. (8) is our main result, connecting vortex entropy to normal conductivity directly.

***Estimations on*** $s_\phi^{\text{amp}}$. We take $\rho_n$ as $\rho_n(T_C)$ for simplicity and use the typical value $\delta\gamma = 2.8$ given by Homes' law [12,13]. The unit of $\rho_n$ is taken to be $\mu\Omega \cdot cm$ as commonly used in experiments, Eq. (8) can be rewritten numerically as:

$$s_\phi^{\text{amp}} = \frac{1}{1+l/\xi} \frac{6.2 \times 10^{-12}}{\rho_n} \mu\Omega \cdot cm \cdot J/(K \cdot m) \quad (8')$$

For dirty ($l/\xi \ll 1$) superconductors with $\rho_n \sim 100\ \mu\Omega \cdot cm$ and c-axis constant of a few nanometers, $s_\phi^{\text{amp}}$ per sheet is around $10^{-23}$ J/K, on the same order of Boltzmann constant $k_B$ [3,7,17]. For general cases, we collect parameters including $\rho_n$, $\xi$ and $l$, calculating $s_\phi^{\text{amp}}$ according to Eq. (8'). The predicted $s_\phi^{\text{amp}}$ are shown in TABLE I, together with the experimentally determined values $s_\phi^{\text{exp}} = N\Phi_0/\rho_{xx}$ at Nernst peaks.

TABLE I. Related parameters, theoretically predicted $s_\phi^{\text{amp}}$ based on Eq. (8'), and experimentally determined $s_\phi^{\text{exp}}$ at Nernst peaks (the corresponding magnetic field and temperature are listed in the brackets) in $FeTe_{0.6}Se_{0.4}$ [18], MoGe [7,19], Nb: $SrTiO_3$ [7], epitaxial $Bi_2Sr_2CaCu_2O_{8+x}$ films (epi-BSCCO) [20,21], exfoliated $Bi_2Sr_2CaCu_2O_{8+x}$ flakes (exf-BSCCO) with different doping levels [9], $YBa_2Cu_3O_{7-\delta}$ [20,22], bulk and thin $NbSe_2$ flakes [23] and $Nb_{80}Mo_{20}$ [24,25].

| Materials | $\rho_n$ [$\mu\Omega$·cm] | $l$ [nm] | $\xi$ [nm] | $s_\phi^{\text{amp}}$ [$10^{-14}$ J K$^{-1}$m$^{-1}$] | $s_\phi^{\text{exp}}$ [$10^{-14}$ J K$^{-1}$m$^{-1}$] |
|---|---|---|---|---|---|
| FeTe$_{0.6}$Se$_{0.4}$ | 400 | 3.4 | 1.6 | 0.50 | 1.6 (24 T, 10 K) |
| MoGe | 150 | 0.3 | 7.7 | 3.9 | 3.1 (4 T, 3.2 K) |
| Nb: SrTiO$_3$ | 140 | 36 | 62 | 2.8 | 2.3 (0.06 T, 0.16 K) |
| epi-BSCCO ($T_C$=90 K) | 140 | 3.4 | 1.42 | 1.3 | 1.6 (12 T, 77 K) |
| exf-BSCCO ($T_C$=90 K) | 180 | 6 | 1.5 | 0.69 | 0.95 (9 T, 75 K) |
| exf-BSCCO ($T_C$=78 K) | 250 | 4 | 1.4 | 0.64 | 0.44 (12 T, 65 K) |
| exf-BSCCO ($T_C$=56 K) | 1000 | 1.8 | 1.6 | 0.29 | 0.053 (12 T, 50 K) |
| exf-BSCCO ($T_C$=36 K) | 4200 | 0.3 | 3.3 | 0.13 | 0.011 (12 T, 35 K) |
| YBa$_2$Cu$_3$O$_{7-\delta}$ | 100 | 4.8 | 1.45 | 1.4 | 1.1 (12 T, 82 K) |
| NbSe$_2$ ($T_C$=4.18 K) | 110 | 11.7 | 9.1 | 2.5 | 3.5 (0.8 T, 2.82 K) |
| NbSe$_2$ ($T_C$=6.13 K) | 7.2 | 106 | 8.1 | 6.1 | 17 (1.2 T, 4.12 K) |
| Nb$_{80}$Mo$_{20}$ | 5.5 | 5.3 | 39 | 99 | 100 (0.03 T, 3.5 K) |

Despite the wide range of $\rho_n$, the difference between $s_\phi^{\text{amp}}$ and $s_\phi^{\text{exp}}$ in each material is within one order. The exceptions are underdoped BSCCO flakes with $T_C \leq 56$ K. One possible reason is that when temperatures are close to $T_C$, the thermal activation of Bogoliubov quasiparticles is significant. While in dirty materials these quasiparticles contribute few Nernst signals due to short mean free path [2], they could produce remarkable $\rho_{xx}$. Therefore, $s_\phi^{\text{exp}} = N\Phi_0/\rho_{xx}$ would be smaller than the mere value of vortex entropy. In fact, in Ref. [9] the temperature of the Nernst peak gradually approaches and exceeds $T_C$ eventually as the doping is decreased.

***Bounds of*** $s_\phi^{\text{amp}}$ ***and*** $N$. In dirty limit ($l/\xi \ll 1$), Eq. (7') leads to a ratio $s_\phi^{\text{amp}}/\sigma_n = 0.7k_B/\sigma_Q \approx k_B \ln 2 /\sigma_Q$ reminiscent of Wiedemann-Franz law [14], implying a particle-transport picture of $2e$ charges carrying entropy and conductivity quanta. It is interesting that $k_B \ln 2$ is the quantum of entropy cost by losing one bit of information, while $\sigma_Q$ is the scale of dissipation per transverse mode in the conductor (differed by a factor 2) according to Landauer's transport formula [26]. The combination of these two scales reminds of Landauer's principle [27] that *erasing information requires energy to be dissipated as heat.* In the present case, the heat should be produced through the quantum dissipator $\sigma_Q$ as each bit of information is erased.

According to boson-vortex duality in 2D disordered superconductors [28,29], superconductor-insulator transition takes place at the critical conductivity $\sigma_c = \sigma_Q$ (the length disappears in units of $\sigma$ in 2D). Therefore, the lower bound to keep superconductivity ($\sigma_n \geq \sigma_Q$) would be:

$$s_\phi^{\text{amp}} \geq \frac{\delta\gamma}{4} k_B \tag{9}$$

If taking $\delta\gamma = 2.8$, then Eq. (9) coincides with the value proposed in Ref. [3], which is reasonable since $k_B \ln 2$ should be the minimum entropy to discern a vortex structure, viz., regions inside or outside the core. If the entropy was smaller, a vortex cannot be defined due to lack of information, and the system would become normal. If the critical conductivity $\sigma_c \neq \sigma_Q$, which, for example, can be tuned in Josephson junction arrays by the ratio of Josephson coupling to Coulomb blockade energy [30], the bound can also be larger or smaller than $k_B \ln 2$. For cleaner materials with $l \geq \xi$, the entropy can also be smaller due to the factor $1/(1 + l/\xi)$ in Eq. (8).

The vortex Nernst signal can be expressed as $N = s_\phi \rho_{xx}/\Phi_0$ [4]. For ideally free vortex liquid, the flux flow resistance [31–33] is $\rho_f = \rho_n H/H_{c2}$, while for pinned liquid, there is also an exponential factor $\exp(-\varepsilon/T)$ characterizing the thermal activation of mobile vortices [34], where the energy barrier $\varepsilon > 0$ depends on pinning details. Therefore, $\rho_{xx} = \rho_f \exp(-\varepsilon/T) < \rho_f$. Considering $1/(1 + l/\xi) < 1$ and $df/d\varepsilon \leq 1$, $N$ follows the inequation:

$$N < \frac{s_\phi \rho_f}{\Phi_0} < \frac{\delta\gamma}{8} \frac{k_B}{e} \left(\frac{3}{2} + \ln\frac{\mu_0 H_{c2}}{B}\right) \frac{B}{\mu_0 H_{c2}} \tag{10}$$

The right-hand side is an increasing function on $B/\mu_0 H_{c2}$, reaching its maximum at $\mu_0 H_{c2}$. However, our result based on London equation neglects the depletion of $n_s$ outside cores due to finite momentum. The inaccuracy could become more significant as the field approaching $H_{c2}$, so we resort to G-L theory at such limit, which takes account of this nonlinear effect. The G-L free energy density [35] is:

$$\begin{aligned}F_{\text{GL}} &= F_n + \frac{B^2}{2\mu_0} - \frac{1}{2\mu_0}\frac{(\mu_0 H_{c2} - B)^2}{1 + (2\kappa^2 - 1)\beta_A} \\ &\sim F_n + \frac{B^2}{2\mu_0} - \frac{1}{2\mu_0}\frac{(\mu_0 H_{c2} - B)^2}{2\kappa^2}\end{aligned} \tag{11}$$

where $\sim$ is associated with G-L parameter $\kappa \sim \lambda/\xi \gg 1$ and the lattice parameter $\beta_A \sim 1$ [35]. However, it should be noted that $F_{\text{GL}}$ contains contributions from both thermally activated Bogoliubov quasiparticles and vortices. The former corresponds to $F_n - \mu_0 H_c^2/2$ under zero field. $F_v$ is extracted from $F_{\text{GL}}$ by subtracting $F_n - \mu_0 H_c^2/2$, as recently used to calculate chemical penitential of vortices [36]. Combined with the relation $H_{c2} = \sqrt{2}\kappa H_c = \Phi_0/2\pi\xi^2$, we obtain $F_v$ of vortices and the corresponding $s_\phi$:

$$F_v = \frac{B^2}{2\mu_0}\left(1 - \frac{1}{2\kappa^2}\right) + \frac{n_v \Phi_0^2}{4\pi\lambda^2}$$

$$s_\phi = \frac{\Phi_0^2}{4\pi\mu_0 \lambda^2(0) T_C} \frac{df}{d\varepsilon} \tag{12}$$

It turns out that Eq. (12) can be extrapolated from Eq. (6) by taking $B = \mu_0 H_{c2}$ and replacing 3/2 with

2. Therefore, the upper bound of $N$ is:

$$N < \frac{\delta\gamma}{4}\frac{k_B}{e} \tag{10'}$$

For $\delta\gamma$ of order unity, Eq. (10') is of order $10\ \mu\text{V/K}$. If taking $\delta\gamma = 2.8$, then $N < k_B \ln 2 /e$.

***The lower bound of viscosity-entropy density ratio***. The viscosity-entropy density ratio $\eta/S$ of vortex liquid was proposed to have a minimum similar with those of classical liquids [3,37]. Here we address this problem. The Bardeen-Stephen (BS) viscosity [31] is $\eta = \mu_0 H_{c2}\Phi_0/\rho_n$. The entropy density is obtained through $S = n_v s_\phi$ after inserting Eq. (8) into Eq. (6). Considering $df/d\varepsilon \leq 1$, the viscosity-entropy density ratio is:

$$\frac{\eta}{S} \geq \frac{8\pi}{\delta\gamma}\frac{\hbar}{k_B}\left[\frac{B}{\mu_0 H_{c2}}\left(\frac{3}{2}+\ln\frac{\mu_0 H_{c2}}{B}\right)\right]^{-1} \tag{13}$$

The right-hand side is a decreasing function on $B/\mu_0 H_{c2}$, reaching its minimum as $B = \mu_0 H_{c2}$. Similar with Eq. (10), the constant 3/2 is replaced by 2 at $H_{c2}$. Therefore, the minimum is, if taking $\delta\gamma = 2.8$:

$$\frac{\eta}{S} \geq \frac{4\pi}{\delta\gamma}\frac{\hbar}{k_B} \sim \frac{\pi\hbar}{k_B \ln 2} \tag{13'}$$

which suggests that if $k_B \ln 2$ sets the scale of entropy, then the viscosity would be set by $\pi\hbar = h/2$, half of the Planck constant. Since BS viscosity is applicable to the free vortex liquid, the general viscosity could be enhanced due to vortex-vortex and vortex-pinning interaction, and Eq. (13) still holds. The lower bound resembles the holographic one in relativistic hydrodynamics [15] where the coefficient of $\hbar/k_B$ is $1/4\pi$, and that of classical liquid [37] which has an additional square root of proton-electron mass ratio.

***Discussion***. The low-field $s_\phi$ by Stephen [5] can be reproduced by replacing the integral cutoff at lattice constant $a$ in Eq. (1) with penetration depth $\lambda$, giving the factor $2\ln\kappa$ rather than $\ln(\mu_0 H_{c2}/B)$. Sergeev's result [6] is the same as our expression for amplitude $s_\phi^{\text{amp}}$. Both results are independent of temperatures, fields, and resistivities. The finite $s_\phi$ close to $H_{c2}$ in Eq. (12) contrasts with that obtained by Maki and Caroli [38–40], where $s_\phi$ should approach zero at $B = \mu_0 H_{c2}$. Their result can be reproduced if calculating $S = -\partial F_{\text{GL}}/\partial T$ directly from Eq. (11), then we would obtain $S \propto (\mu_0 H_{c2} - B)$. The discrepancy can be resolved by noting that $S$ calculated from $F_{\text{GL}}$ also contains the contribution from thermally activated quasiparticles. To estimate vortex entropy, we have utilized Homes' law to deduce $\delta\gamma$. The procedure can also be reversed, for example, one obtains $s_\phi$, the parameter $\gamma$ and characteristic length scales like $\xi$ and $l$ through thermoelectric transport and scanning-tunneling microscopy, the superconducting portion $\delta$ can then be estimated.

In summary, we have derived the average entropy $s_\phi$ of each vortex through a thermodynamic method. We further relate its amplitude $s_\phi^{\text{amp}}$ with the normal-state resistivity. The consistency with experimental values suggests our consideration is reasonable. The inclusion of resistivity also enables us to estimate bounds to vortex entropy, Nernst signals and viscosity-entropy density ratio. The factor $\ln 2$ is incorporated from $\delta\gamma = 2.8$ given empirically by Homes' law [12,13]. Therefore, the last piece of the whole picture might reside in the mystery of Homes' law.

***Acknowledgment.*** We are grateful to discussions with K. Behnia, G. E. Volovik, Ding Zhang, and Yang Wang. The work was supported by the National Key Research and Development Program of China (2021YFA0718800).